\documentclass[11pt]{article}

\usepackage[utf8]{inputenc}
\usepackage[T1]{fontenc}
\usepackage{lmodern}
\usepackage{textcomp}

\usepackage[letterpaper,margin=1in]{geometry}
\usepackage{setspace}
\usepackage{authblk}

\usepackage[hidelinks]{hyperref}
\usepackage{xurl}

\title{\bfseries Care, Human Enfeeblement, and the Existential Implications of AGI}

\author[1,2]{B. Scot Rousse\thanks{ORCID: 0000-0002-6840-9641. Email: \texttt{bsrousse@berkeley.edu}}}
\affil[1]{UC Berkeley, Berkeley CA 94720, USA}
\affil[2]{Topos Institute, Berkeley CA 94704, USA}
\date{}

\begin{document}

\maketitle

{\renewcommand{\thefootnote}{}\footnotetext{M.~Ikl\'e et al. (Eds.): AGI 2026, LNAI 16855, pp.~211--218, 2026.\\ \url{https://doi.org/10.1007/978-3-032-33195-3_16}}}

\begin{abstract}
\noindent Stuart Russell has named ``human enfeeblement'' as a long-term societal risk of advanced AI. This paper gives a new interpretation of the enfeeblement danger by interpreting it in light of recent work by Brian Christian, Alison Gopnik, and David Spivak on care and caregiving, and connecting it with the ontological design tradition of Terry Winograd and Fernando Flores. The enfeeblement that threatens is not merely a matter of cognitive off-loading and loss of knowledge, but a form of disengagement with the world which I explain in terms of an erosion of our capacities to care. This paper articulates a conception of care as a constellation of cultivable skills that involve us in the world and with each other, while making life worth living. Designing AI to support our capacities to care addresses the worry recently raised by Demis Hassabis and others regarding the questions of ``meaning and purpose'' that are looming in a post-AGI world.

\end{abstract}

\section{Care, Enfeeblement, and the Challenge of AGI}

A growing number of scholars have been arguing that we should place \emph{care} at the center of how we think about AI and the implications of AGI. In his AGI-2024 Keynote, David Spivak \cite{spivak2024} argued that Turing's foundational model of computation omitted the element of care that motivated the human ``computers'' his model ``biomimicked.'' Alison Gopnik's research \cite{gopnik2023daedalus,gopnik2023aps} reveals an intelligence intrinsic to caregiving that current AI paradigms miss. Brian Christian \cite{christian2025} traces how AI's computational frameworks have been evolving toward a caregiving relationship with users, using this progression to sharpen the question of what genuine care requires and where current systems fall short. And Terry Winograd \cite{winograd2024} insists that designers of AI technologies must account for ``the role of human care and concern.'' This paper takes up and consolidates this rising tide of interest in care, using it to flesh out the stakes of what Stuart Russell has identified as a disconcerting long-term societal risk of advanced AI: human enfeeblement \cite{russell2019}.

Russell's enfeeblement names the danger that, as AI systems take over more and more human work, we will gradually lose the knowledge and know-how required to maintain our own civilization, becoming ``passengers on a cruise ship run by machines'' \cite[p.~255]{russell2019}. He rightly insists that this requires a cultural, not just a technical, response. His sketch of a remedy---the cultivation of a cultural ethos of ``autonomy, agency, and ability''---is a helpful start, but it misidentifies what is most at risk. The real threat of enfeeblement is a matter not just of a loss of knowledge or autonomy, but a dilution of our meaningful engagement in the world and with each other. I will spell out this existential disengagement in terms of an erosion of our \emph{capacities to care}. To care is to be drawn into responsive involvement with each other and the world, to tend to and articulate what matters, to stake commitments under uncertainty, and to take care of shared concerns together. This reorientation has deep and varied roots. Nel Noddings \cite{noddings2013} and Patricia Benner \cite{bennerwrubel1989} established care as a fundamental human orientation, not merely a sentiment. John Haugeland noted the omission Spivak identified in Turing when he observed, drawing on Heidegger \cite{heidegger1962}, that ``the trouble with artificial intelligence is that computers don't give a damn'' \cite{haugeland1979}. My contribution here is to bring this phenomenological and care-centered critique into the context of AGI implications by interpreting enfeeblement as an erosion of the capacities through which human beings find, articulate, commit to, and coordinate around what matters. Care---understood not as a feeling of affection but as a constellation of cultivable skills---is what is most at stake in the development of AGI.

\section{From Caregiving to Caring as Such}

Gopnik's ongoing work \cite{gopnik2025} demonstrates that human caregiving is at the neglected center of so-called ``human-level intelligence.'' A good caregiver acts not to produce a predetermined outcome but to support the cared-for person's own developing, autonomous agency. Gopnik \cite{gopnik2023aps} convincingly stresses the relevance of this intelligence for how we train AI. Christian \cite{christian2025} takes up this point. He shows that AI alignment frameworks have been evolving toward a caregiving relationship, moving from hand-coded behavior to reward design, to learning from demonstrations, to learning from human preference judgments. He argues that this approaches a caregiving relationship in two senses. First, in the sense that training AI is similar to raising a child in that it aims to produce an agent which can act on its own and adapt to a changing world; and second, in aiming to produce systems that will eventually behave as caretakers toward us, once AI becomes the more capable party and the current power asymmetry reverses.

Christian sees the kind of enfeeblement imagined by Russell as a possible effect of a distorted, passivity-inducing caregiving relation that powerful AI systems may eventually take toward us (``spoiling us'' as it were) unless we can figure out how to better teach our AI systems about care. In his essay Christian sketches some elements that a non-pathological, true caregiving relation would involve. For example, he stresses the crucial point that a caregiver often acts against the cared-for's immediate revealed preferences in order to promote their overall flourishing, caring for not just what they ``want'' but for what is ``good'' for them. As he puts it, ``as carers, we aspire to serve the `higher' purposes'' \cite[p.~192]{christian2025}.

A caregiving relation that aims to develop the cared-for into an autonomous agent capable of attuning to and acting on ``higher'' purposes requires helping them develop the capacities to ``give a damn'' in Haugeland's \cite{haugeland1979} blunt phrase. In brief, caregiving aims to develop agents who themselves are capable of caring and being meaningfully engaged in life. This means helping them develop the capacity to be motivated not only by what they want (their immediate, revealed preferences), but what they ``want to want,'' what Frankfurt \cite{frankfurt1988} calls their ``higher order'' desires. In the case of human agents, these higher order desires pertain to what we love and care about, what we are motivated to tend and attend to, the kind of person we aspire to be, the kind of world we are driven to create. To care is what enables us to engage in design in Simon's \cite{simon1996} sense---to apprehend an ``existing situation'' as failing to meet our normative standards, and to orient ourselves toward a better, ``preferred one.'' It is this capacity to ``give a damn'' that motivates human agents to be engaged in the world and with others, and to strive toward a life and a world that more closely aligns with their sense of what is good and worthwhile. In this picture, the ``human values'' at issue in alignment research are not fixed objects that exist prior to care, waiting to be elicited and encoded. Values are grounded in care: generated and realized through the practices by which things come to matter to us, are articulated in language, tested in commitment, and enacted in coordinated action.

The human caregiving relationship itself is thus oriented to a broader phenomenon: what we might call \emph{caring as such}.\footnote{The concept of \emph{caring as such} developed here takes up Heidegger's \cite{heidegger1962} analysis of \emph{Sorge} (care) as indicating the fundamental character of human life. In doing so, this paper also draws on and extends the tradition of phenomenology-influenced technology critique that runs from Dreyfus \cite{dreyfus1992} through Haugeland \cite{haugeland1979}, Winograd and Flores \cite{winogradflores1986}, and Smith \cite{smith2019}. Borgmann's account of the ``device paradigm'' and its tendency to displace the focal practices through which we sustain what matters \cite{borgmann1984}, along with Turkle's work on how digital mediation erodes the relational capacities that caring requires \cite{turkle2011,turkle2015}, are further important adjacent voices in this tradition.} In the words of the nurses, Benner and Wrubel: ``Caring \ldots{} means that persons, events, projects, and things matter to people. Caring is essential if the person is to live in a differentiated world where some things really matter while others are less important or not important at all'' \cite[p.~1]{bennerwrubel1989}. It is important to stress that caring in this sense is morally neutral. Drug cartels care about their operation and find a meaningful orientation in the world through it just as much as PETA or M\'edecins Sans Fronti\`eres do. This moral neutrality is why care demands our reflection.

Care is the capacity through which persons, projects, and relationships are lived as mattering at all; it names the structure of engaged involvement, not yet the moral quality of that involvement. Further constraints are needed for that engagement to count as ethically responsive. In general terms, ethical care also requires respecting constraints around accountability, autonomy, reciprocity, and non-domination, so that what matters to one person or group remains answerable to the claims of others. Caring responsibly means caring about the ethical dimensions of what we care about. Much more needs to be said about this topic, especially since the emergence of AGI itself will likely scramble and challenge our inherited moral intuitions and criteria for moral standing \cite{rousse2026aaai}.

In sum, care is our fundamental orientation, the way we find ourselves involved with others in a world that makes a difference to us, wherein we find projects and relationships \emph{worth} pursuing and devoting ourselves to. If this orientation erodes, the world becomes less differentiated---less full of things that matter, leaving us passive, uninterested, and existentially \emph{enfeebled}. It is imperative that a sensibility for care guide the design of AI systems.

\section{Ontological Design and the Skills of Care}

The claim that AI research needs a richer picture of human agency has a notable lineage within the AI field itself. In 1986, Winograd and Flores published \emph{Understanding Computers and Cognition} \cite{winogradflores1986}, challenging the then-dominant ``rationalistic'' picture of language and intelligence in AI and proposing an alternative grounded in phenomenological philosophy, biology, and Speech Act Theory. One of their central insights was what they called ``ontological design'': in designing computational tools, we are not merely solving problems; we are designing ways of being. Every technology embeds assumptions about what human beings are and how they should relate to the world and to each other. Therefore, the designers of computer systems need to be aware of the ontological implications and presuppositions of their work. In this vein, Winograd and Flores articulated an ontology of language that defined it as the ability to coordinate commitments in order to take care of concerns, rather than just as a medium to transmit neutral information, and they offered this as part of their richer foundation for the design of computer systems. Today, we need an ontology of care to guide the design of AI systems because AGI intensifies the problem of ontological design: as systems become more agentic, conversational, personalized, and cross-domain, they will increasingly mediate and possibly distort our way of being. If care is a capacity that is at risk and whose preservation can help set design constraints for how we build and deploy AI systems, we need to be specific about what it consists in, again, not as a sentiment of affection, but as a set of cultivable skills. Let me briefly sketch four here in broad strokes.

\emph{\textbf{Receptivity}} is the capacity to be moved by emerging concerns that call to us as important, worthy of our attention \cite{frankfurt1988,spinosa1997}. What we care about is something we find ourselves drawn to---what solicits our involvement before we have calculated its utility, the way that \emph{falling} in love precludes utility calculation. An AI system that resolves every ambiguity, smooths every friction, and pre-configures a menu of options (like today's dating apps or online shopping behemoths) works against this capacity to be \emph{struck by} and \emph{fall for} what one encounters. Designing for receptivity means resisting the logic of seamless convenience while leaving an opening for meaningful contingencies and a sense for what does not fit our pre-arranged plans.

\emph{\textbf{Linguistic articulation}} is the skill of focusing an emerging concern into communicable form---transforming a vague sense that ``something is off'' into a commitment one can share, sustain, and coordinate around over time \cite{denningrousse2024,taylor2016}. For example, Arlie Hochschild's phrase ``emotional labor'' \cite{hochschild1983} made salient the management of feeling and emotional display that workers perform as part of paid service work, especially in gendered, traditionally ``feminized'' roles. Miranda Fricker shows the dynamic of articulation at work in the coining of the phrase ``sexual harassment'' \cite{fricker2007}, which helped women bring into better focus and organize against pervasive mistreatment they experienced at work. Russell's notion of ``human enfeeblement'' is another act of linguistic articulation that turned an unease about AI-induced heteronomy into a recognizable concern. To design for linguistic articulation is to support users in focusing their own emerging concerns into provisional, communicable form, rather than generating that articulation on their behalf. If we allow AI to draft our communications, positions, and arguments for us, we sever the fecund movement from inarticulate concern to genuine commitment.

\emph{\textbf{Commitment under uncertainty}} is the gumption required to stake oneself on a project, concern, or relationship despite the absence of guarantees. Frankfurt \cite{frankfurt1988} pointed to Martin Luther's declaration, ``Here I stand, I can do no other,'' as an example of how deep care can compel action that transcends purely rational calculation and flies in the face of established convention, undaunted by uncertainty. Yet the uncertainty at stake pertains not only to consequences. It also includes uncertainty about how our own cares, commitments, and self-understanding may change as we live them out. An AI system that promises to eliminate uncertainty, rigidly align us with our past avowals, always recommend the ``best'' option, minimize risk, and optimize every outcome risks training users out of the capacity to take meaningful risks.

Designing for commitment under uncertainty means helping users clarify what matters and what is at stake, imagine tradeoffs, and anticipate consequences without needing to remove ambiguity or make things totally safe and predictable. Such systems should also be able to support the difficult human art of recommitment and reorientation: helping users keep faith with the stands they have taken while preserving the courage to revise or release them when care itself calls for transformation.

Finally, caring effectively requires the conversational skills that enable \emph{\textbf{conscientious coordination}} of commitments. As Flores \cite{flores2013} and Winograd and Flores \cite{winogradflores1986} showed, human enterprises and institutions are constituted as a network of conversations through which we coordinate the commitments generated by requests, promises, and assertions, in order to take care of concerns. Our shared endeavors flourish when there is a culture of honoring this commitment-dimension of language, going beyond a picture of language as a medium to transmit information. This is, in practice, a concrete realization and refinement of what Spivak \cite{spivak2024} called the ``fabric of belonging'': an ongoing, active tending to the web of mutual relationality through which we make sense together and tend to what matters in common.

What is at stake in the design of advanced AI systems is nothing less than the conditions under which this fabric of belonging can be sustained. The danger is that AI systems that handle scheduling, management, communication, and coordination on our behalf may allow the practices through which responsibility is maintained and trust is built to atrophy: requesting, promising, assessing, repairing breakdowns, and holding one another mutually accountable. This would amount to what Jonker \cite{jonker2024} calls the ``thinning'' of an organization's ``cooperative interface.'' Winograd and Flores's \cite{winogradflores1986} account of language as the medium through which we coordinate our care for shared concerns helps us see why AI agents pose a distinctive risk when they take over the conversational work through which communicative competence is cultivated and exercised \cite{rousse2026expertise}.

All four of these skills share a crucial temporal feature: they are always outstanding. As long as you care about something, there is always more to do, more ways that it might flourish or fail. This open-endedness is the opposite of a solved problem, and it contributes to care's essential fragility: these capacities must be practiced in order for our care to be sustained. This is why the risk posed by optimization-driven AI is not merely substitutive but corrosive. When a system collapses every ambiguity and pre-selects the ``best'' option, it forecloses the very occasions on which the skills of care are practiced and sustained.\footnote{The design reorientation suggested here resonates with HCI traditions that have explored how to design for reflection, plurality, and enriched experience, often by resisting pure, seamless interaction: from McCarthy and Wright's interpretation of technology in terms of felt experience \cite{mccarthywright2004} to Sengers et al.'s work on reflective design \cite{sengers2005}, to Bardzell's feminist HCI \cite{bardzell2010}.}

\section{Conclusion: Towards a Design for Care}

Winograd and Flores's \cite{winogradflores1986} principle---that in designing tools we are designing ways of being---has never been more urgent. If the human capacity to care atrophies, we do not simply become less capable; we become disengaged, existentially enfeebled beings to whom nothing really matters. These are the true stakes of Russell's problem of enfeeblement. The practical upshot is a design reorientation: from catering to caring. The aim, then, goes beyond aligning AI with human values that are already in hand. Our task is to ensure that our AI systems support the human practices and processes through which values are generated and realized. Current AI systems overwhelmingly cater: they serve up what is wanted, optimize for revealed preferences or pre-given values, and satisfy the immediate request. If we leave this design practice in place, we will create systems that support ease and leisure but not meaning. This elucidates Demis Hassabis's recent, recurrent emphasis on the ``question of meaning and purpose'' in a post-AGI world \cite{hassabis2025}. Designing for care means taking the question of meaning and purpose seriously and doing so concretely by orienting to the four cultivable skills discussed above. These are the capacities from which flow human value and meaningful engagement in the world.

Taken together, these four skills provide a preliminary framework for the design of AI systems. To evaluate a system in terms of care, we can ask whether it expands or narrows receptivity to what matters, supports or replaces the user's own work of articulation, clarifies commitment under uncertainty or absorbs that commitment into recommendation, and strengthens human coordination or bypasses the practices of request, promise, assessment, repair, and accountability. AI becomes enfeebling when it removes the occasions through which these capacities are practiced and engaging when it scaffolds their continued exercise.

The systems now emerging from the world's leading AI laboratories are reconfiguring the conditions under which human beings learn to care and engage with each other and the world \cite{rousse2026expertise}. The question for AGI 2026 is whether we can design tomorrow's AI systems in light of the fact that we are beings whose quality of life is defined by meaningful engagement with what we care about. I hope that other thinkers, engineers, and designers will join the project of imagining and implementing what this could amount to in practice.

\subsection*{Acknowledgments}
My articulation of several themes above draws upon forthcoming collaborative work with David Spivak. I give thanks to Fernando Flores, Marjan Sharifi, David Spivak, Terry Winograd, and Alex Zhu for conversations that shaped this paper. Three anonymous referees for the AGI-26 conference also provided useful and detailed feedback on an earlier draft.

\subsection*{Disclosure of Interests}
The author has no competing interests to declare that are relevant to the content of this article.

\end{document}